\documentclass[final,3p,times,twocolumn]{elsarticle}

\usepackage{amssymb}
\usepackage{amsmath}
\usepackage{lipsum}
\usepackage{color}
\usepackage[none]{hyphenat}
\usepackage{hyperref}  
\usepackage{multirow}

\biboptions{numbers,sort&compress}

\journal{CPC}

\begin{document}

\newcommand{\ud}{\mathrm{d}}
\newcommand{\ui}{\mathrm{i}}
\newcommand{\ue}{\mathrm{e}}
\newcommand{\MeV}{\text{MeV}}
\newcommand{\cm}{\text{cm}}
\newcommand{\K}{$^\text{40}$K}
\newcommand{\U}{$^\text{238}$U}
\newcommand{\Th}{$^\text{232}$Th}

\begin{frontmatter}



\title{Hunting the Potassium Geoneutrinos with Liquid Scintillator Cherenkov Neutrino Detectors}

\author[THU,CHEP]{Zhe~Wang\corref{cor1}}
\author[THU,CHEP]{Shaomin~Chen}

\cortext[cor1]{Email: wangzhe-hep@mail.tsinghua.edu.cn}

\address[THU]{Department~of~Engineering~Physics, Tsinghua~University, Beijing 100084, China}
\address[CHEP]{Center for High Energy Physics, Tsinghua~University, Beijing 100084, China}

\begin{abstract}
The research of geoneutrino is a new interdisciplinary subject of particle experiments and geo-science.
Potassium-40 (\K) decays contribute roughly 1/3 of the radiogenic heat of the Earth, but it is still missing from the experimental observation. Solar neutrino experiments with liquid scintillators have observed uranium and thorium geoneutrinos and are the most promising in the low-background neutrino detection.
In this article, we present the new concept of using liquid-scintillator Cherenkov detectors to detect the neutrino-electron elastic scattering process of \K\ geoneutrinos.
Liquid-scintillator Cherenkov detectors using a slow liquid scintillator can achieve this goal with both energy and direction measurements for charged particles. Given the directionality, we can significantly suppress the dominant intrinsic background originating from solar neutrinos in conventional liquid-scintillator detectors. We simulated the solar- and geo-neutrino scatterings in the slow liquid scintillator detector, and implemented energy and directional reconstructions for the recoiling electrons. We found that \K\ geoneutrinos can be detected with three standard deviation accuracy in a kiloton-scale detector.
\end{abstract}

\begin{keyword}
Liquid-scintillator Cherenkov detector \sep Slow liquid scintillator \sep Geoneutrino \sep $^{40}$K neutrino


\end{keyword}

\end{frontmatter}


The interaction of neutrinos with matter is extremely weak, so they easily penetrate celestial bodies.
Determining the neutrino spectrum and flavor can shed light on the production reaction and environment.
They are thus ideal probes of the Earth and Sun.
Geoneutrinos are primarily generated by three types of long-lived radioactive isotopes, potassium-40 (\K), uranium-238 (\U), and thorium-232 (\Th).
Their origins, compositions, and distributions are very interesting questions in geoscience.
They can help to discover the physical and chemical structure of the Earth and even to reveal its evolution.

The KamLAND~\cite{KamLAND, KamLAND2, KamLAND3, KamLAND4} and Borexino~\cite{BX, BX2, BX3} experiments have made pioneering contributions to geoneutrino discovery.
The detection is achieved by finding inverse-beta-decay, IBD, signals in liquid-scintillator detectors.
An IBD signal consists of a prompt positron signal and a delayed neutron-capture signal,
and the prompt-delay-coincidence provides a clear signature of the interaction. The IBD cross-section is relatively high.
The energy threshold for the reaction is 1.8 MeV, and only \Th\ and \U\ geoneutrinos are accessible.
Almost no directional information can be extracted for the initial neutrinos~\cite{JohnIBD}.

Neutrinos originating in the mantle have a direct connection with
the power that drives the plate tectonics and mantle convection~\cite{Mantle1,Mantle2}.
However, the measurements of mantle geoneutrinos rely heavily on the crust geoneutrino predictions.
Consequently, the mantle component still has considerable uncertainty.
In~\cite{Li6LS}, Tanaka and Watanabe proposed to use a $^\text{6}$Li-load liquid scintillator to extract directional information about U and Th neutrinos using the IBD process, and even to image the Earth's interior.

The K element is more mysterious than U and Th.
Potassium-40 (\K) decays contribute roughly 1/3 of the radiogenic heat of the Earth, but no experimental \K\ neutrino result has been reported.  And because K is a volatile element, precipitating it into typical mineral phases is more difficult than for U or Th.
Measuring the flux of \K\ neutrinos versus \U\ and \Th\ neutrinos can offer input to understanding the formation process of the Earth~\cite{K40Mantle2}.
The model of \K\ and $^\text{40}$Ar in the air and the Earth also indicates the enriched and depleted mantle structure~\cite{K40Mantle1}.
In~\cite{GasTPC}, Leyton, Dye, and Monroe proposed to use directional neutrino detectors, like noble-gas time-projection chambers to explore various geoneutrino components.

Note that, so far, the liquid scintillator detector of Borexino is the only one achieving sub-MeV neutrino spectroscopy, where
both large target mass and low background are realized.
Solar neutrinos are detected through neutrino-electron scattering,

\begin{equation}
\nu+e^- \rightarrow \nu+e^-,
\label{eq:nueScat}
\end{equation}
which has no theoretical threshold. There is a strong correlation between the initial neutrino and scattered electron direction, especially after imposing a requirement on the kinetic energy of the recoiling electron.
In this paper, we consider introducing directionality to the conventional liquid scintillator detector
to suppress the intrinsic solar neutrino background and to detect the \K\ geoneutrinos.

Because of the long emission time constant of scintillation radiation, the new type of liquid-scintillator Cherenkov neutrino detectors~\cite{JohnLearned, YEH201151, THEIA} can identify the small prompt Cherenkov radiation from huge slow scintillation light.  This unique feature can provide not only the reconstruction of both direction and energy which has never been achieved in conventional liquid scintillation neutrino detectors but also particle identification~\cite{Hanyu}.
We consider two schemes for the liquid-scintillator
Cherenkov neutrino detector. The first approach is to use a fast, high light-yield liquid scintillator and fast photon detectors. In this case,  the light yield can reach 10,000 photons per MeV with an emission time constant of a few nanoseconds, which is much longer than several picoseconds of timing precision of the photon detectors. The recent experimental development for this approach can be found in references~\cite{CHESS,CHESS2,LAPPD}. The second approach is to use a slow liquid scintillator and photomultiplier tubes (PMTs). The PMTs usually have a timing precision of about one nanosecond, while the emission time constant for the slow liquid scintillator is much longer, for example, 20 nanoseconds~\cite{SlowLS, mohan, Gruszko:2018gzr}.

\begin{figure}[h]
\centering
\includegraphics[width=0.45\textwidth]{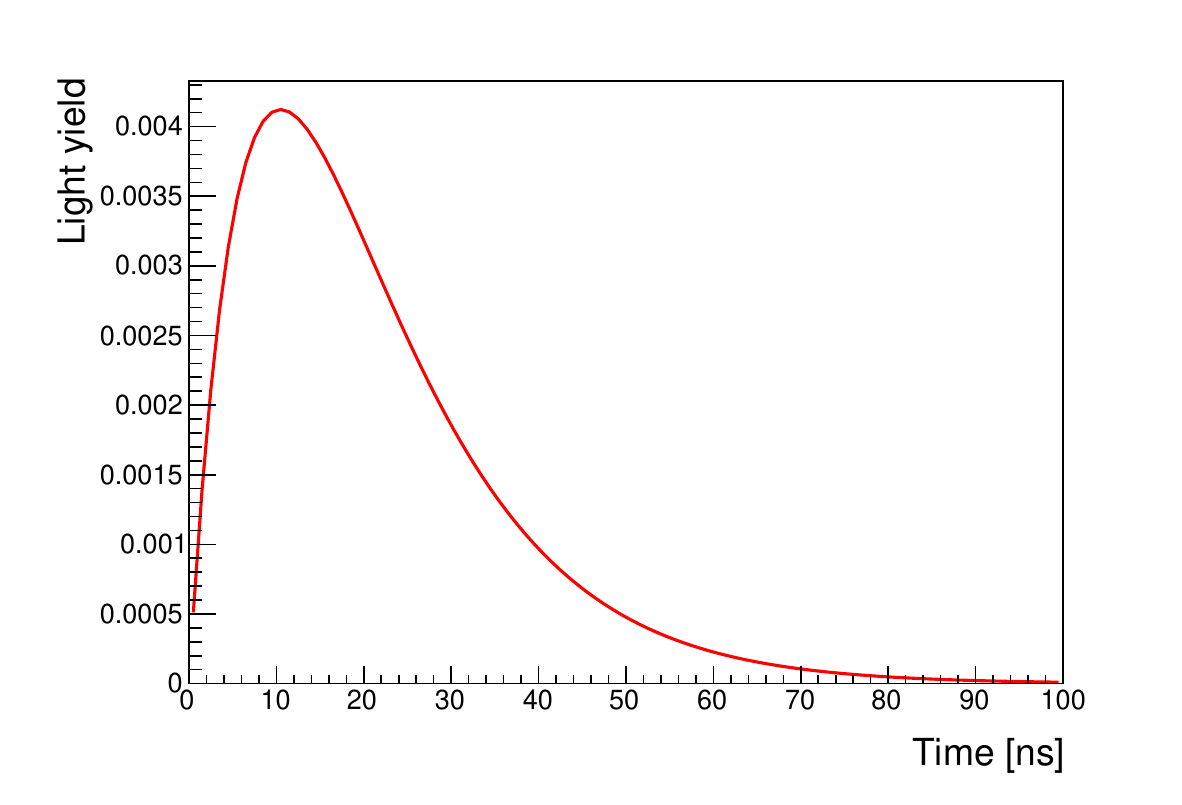}
\caption{The normalized time profile of scintillation light emitted by the slow liquid scintillator, LAB~\cite{SlowLS}.}
\label{fig:TimeProfile}
\end{figure}

In this article, we focus on the latter scheme.
We adopted the parameter set for pure linear alkylbenzene (LAB)~\cite{SlowLS} as a slow liquid-scintillator candidate,
which is most favorable for Cherenkov separation.
The light yield is 2530 photons/MeV, and the emission rise and decay time constants are 12.2 ns and 35.4 ns, respectively.
The time profile is shown in Figure~\ref{fig:TimeProfile}.
The Cherenkov threshold is 0.178 MeV, assuming the refractive index of the liquid to be 1.49~\cite{index}.

\section{Analysis and Result}
\label{sec:results}

\subsection{Ideal Expectation with a Terrestrial Detector}
\label{sec:principle}
\begin{figure}[h]
\centering
\includegraphics[width=0.4\textwidth]{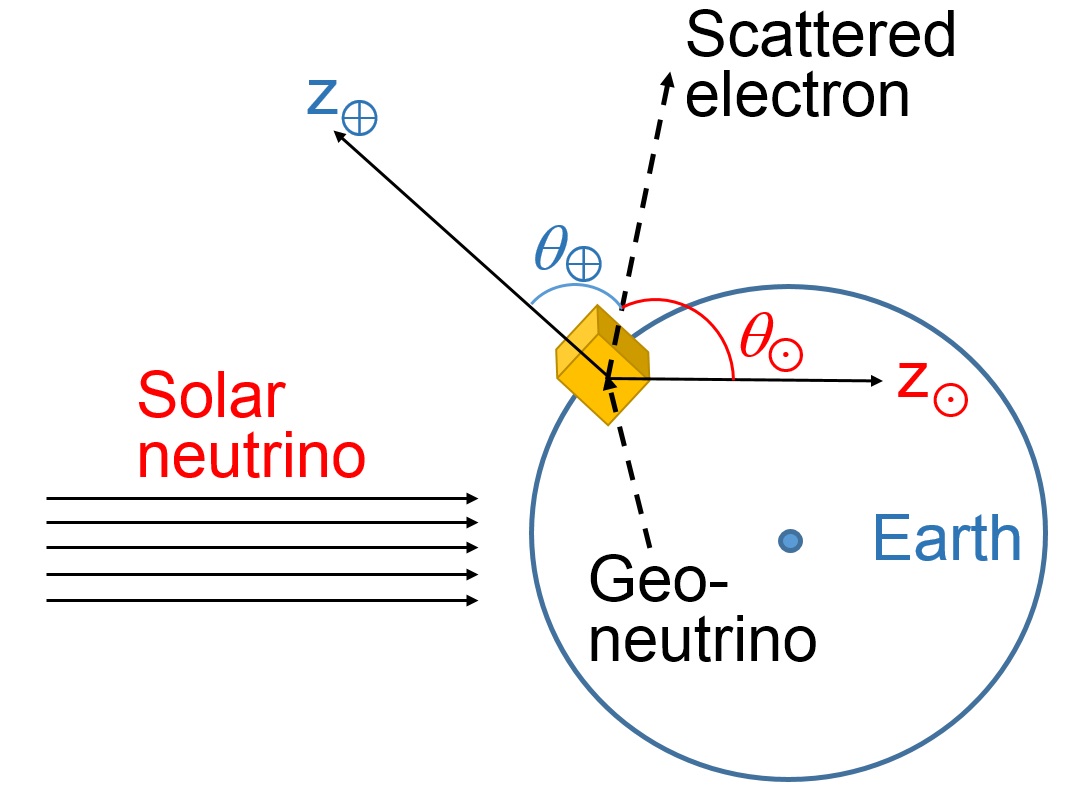}
\caption{Definitions of $z_{\odot}$, $z_{\oplus}$, $\theta_{\odot}$, and $\theta_{\oplus}$. The yellow cube represents the neutrino detector.}
\label{fig:defination}
\end{figure}

\begin{figure}[h]
\centering
\includegraphics[width=0.45\textwidth]{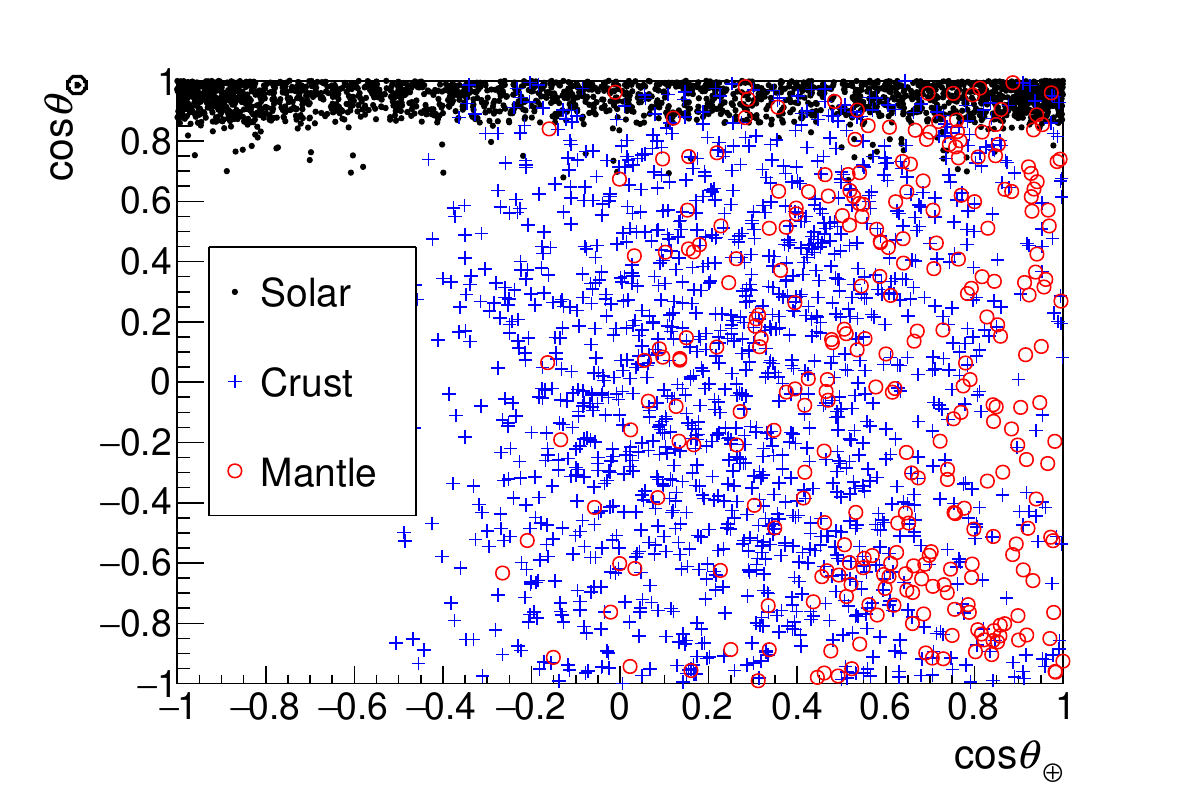}
\caption{Theoretical distributions of $\cos\theta_{\odot}$ and $\cos\theta_{\oplus}$ for solar, crust and mantle neutrinos when the kinetic energies of the recoiling electrons are required to exceed 0.7 MeV.}
\label{fig:Sun_Ear}
\end{figure}

We consider an ideal terrestrial detector, located on the Earth's equator, rotates along with the Earth.
We define solar z-axis, $z_{\odot}$, from the Sun to the Earth, and an Earth z-axis, $z_{\oplus}$, from the center of the Earth to the detector
(see Figure~\ref{fig:defination}). Correspondingly,
we define the angle between the recoiling electron and $z_{\odot}$ as $\theta_{\odot}$ and the angle with $z_{\oplus}$ as $\theta_{\oplus}$.
Geoneutrinos and solar neutrinos are generated and the kinetic energies of the recoiling electrons are recorded.
With a cut on the recoiling electron kinetic energy at 0.7 MeV, the distributions of $\cos\theta_{\odot}$ and $\cos\theta_{\oplus}$ for the remaining neutrinos are shown in Figure~\ref{fig:Sun_Ear}.
The geoneutrinos (crust and mantle) are clearly separated from the solar-neutrino background.
Since the energy range of the \K\ neutrinos is distinguishable from those of the \Th\ and \U\ neutrinos, there is a chance to detect the \K\ component. Next, we will consider a real detector.


\subsection{Liquid-Scintillator Cherenkov Detector Simulation}
\label{sec:Detector}
\begin{figure}[!h]
\centering
\includegraphics[width=0.35\textwidth]{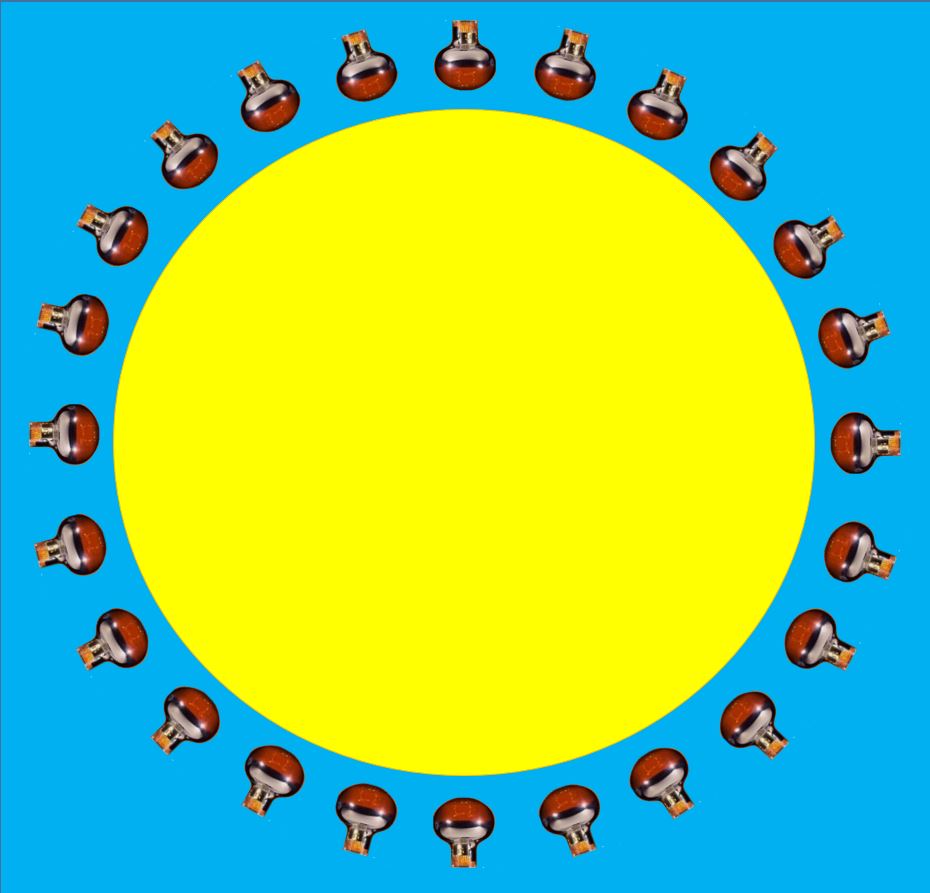}
\caption{Detector concept. From inside to outside are the slow liquid scintillator (yellow), water (blue), PMT, and water again.}
\label{fig:Detector}
\end{figure}
We adopted the detector scheme~\cite{BXDetector} shown in Figure~\ref{fig:Detector}.
In the center is the slow liquid scintillator, which is contained in a transparent container, and it is surrounded by a non-scintillating material, such as water or mineral oil. The PMTs are installed with all photocathodes facing inward and forming a large spherical array. The PMTs are all immersed in the water or oil. The water behind the PMTs also serves as a veto layer, shielding the detector from radioactivities like betas, gammas, neutrons, and cosmic-ray muons.
The number of signals is directly proportional to the target mass,
and due to the required low-background rate, only the central region of the liquid scintillator known as the fiducial volume is accepted for signal detection.

The simulation of solar-neutrino generation, geo-neutrino generation, and neutrino-electron scattering are described in \ref{apx:Solar}, \ref{apx:Geo}, and \ref{apx:Scat}, respectively.
The recoil electrons are simulated using Geant4~\cite{G4-1, G4-2, G4-3} including all possible electromagnetic processes.
Because of multiple scattering, the initial direction of the electron is smeared out.
Some electrons eventually turn back, when they are close to stopping.

Both Cherenkov and scintillation light emissions are handled by Geant4;
however, the production of scintillation light is customized according to LAB measurement~\cite{SlowLS}.

All the optical photons are recorded and gone through empirical simulations~\cite{BXDetector, Logan},
because the attenuation length of optical photons still needs more experimental research~\cite{SlowLS}, and
the target mass or volume is a parameter we want to test.
The limited PMT photocathode coverage and photon attenuation and scattering will cause efficiency loss, so we assume practically only 66.7\% (2/3) of the photons can reach the PMTs.
The quantum efficiency of a PMT is assumed to be 30\% for all photons within the range [300, 550] nm and to be zero for the rest, which is motivated by the high quantum efficiency of PMTs, according to~\cite{hamamatsu}.
In summary, the total efficiency for generating photoelectron, PE, is 20\% for photons within the range [300, 550] nm and is zero outside of this wavelength range.

\subsection{Energy and Direction Reconstruction}
\begin{figure}[!h]
\centering
\includegraphics[width=0.5\textwidth]{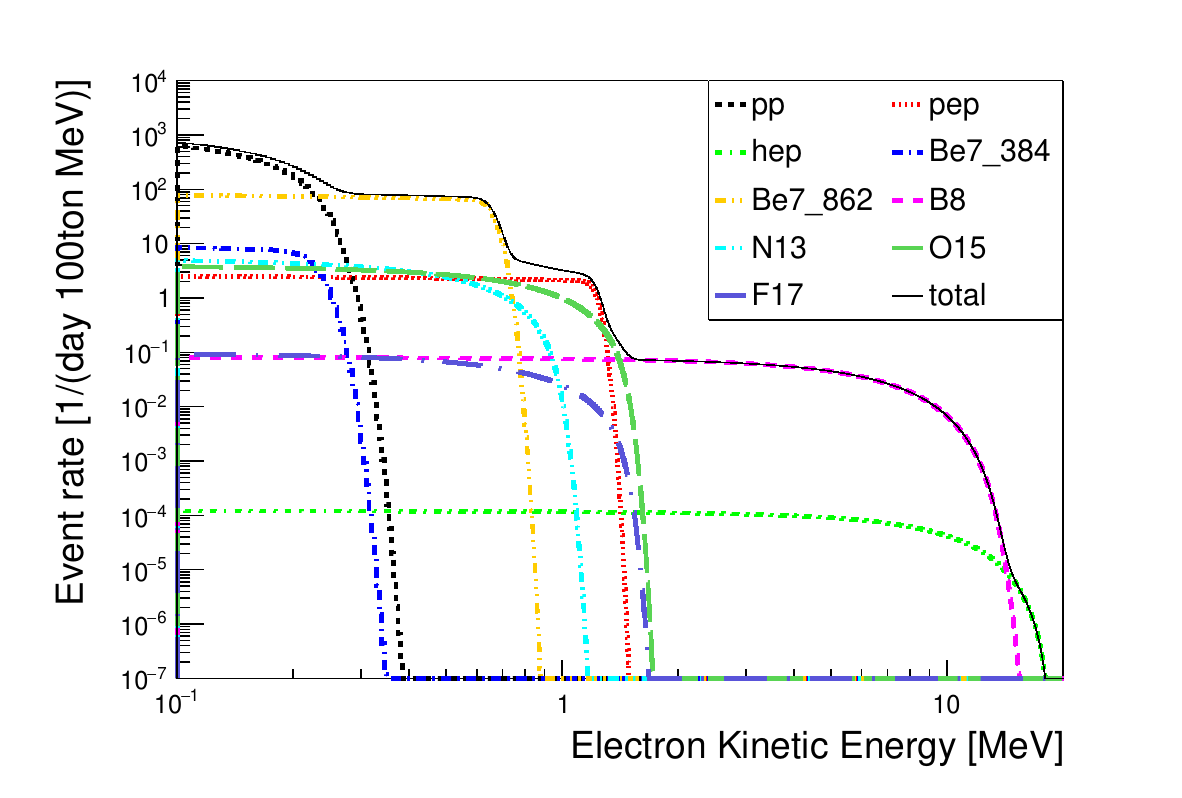}
\caption{Detected kinetic-energy spectra of the recoiling electrons produced by solar neutrinos in the simulation.}
\label{fig:SolarDet}
\includegraphics[width=0.5\textwidth]{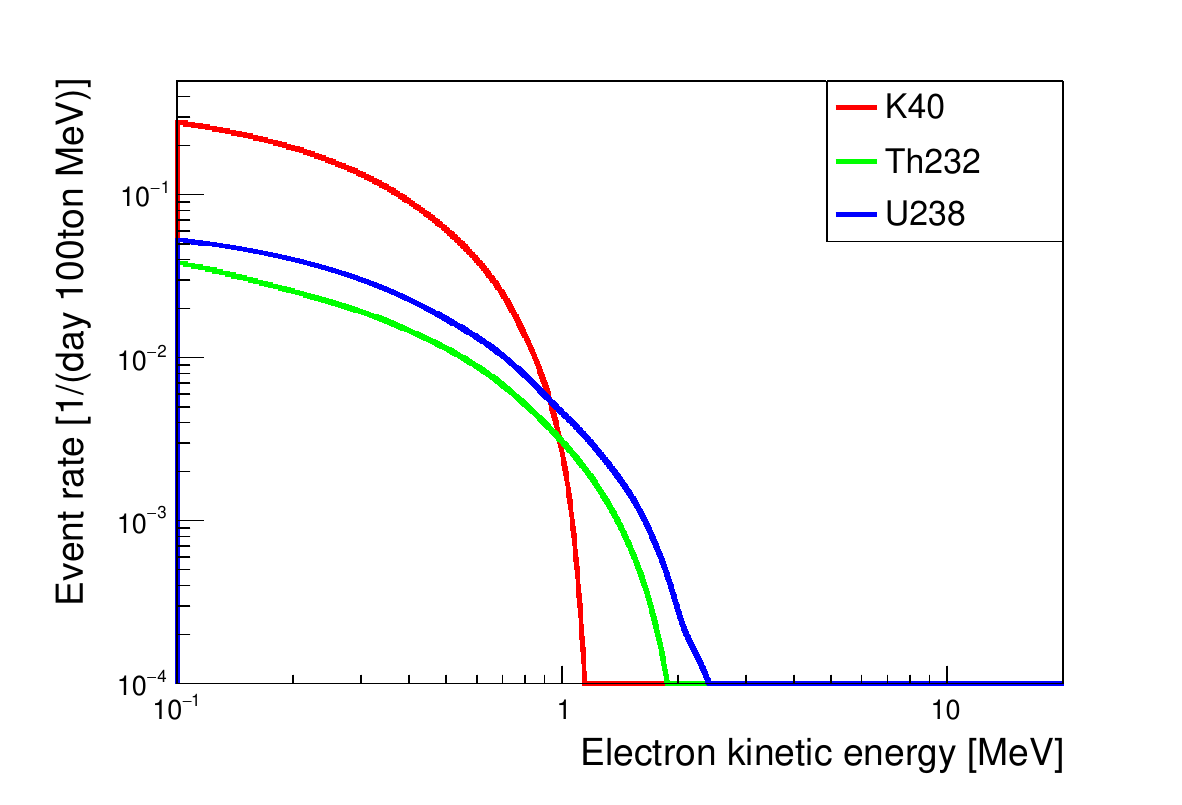}
\caption{Detected kinetic-energy spectra of the recoiling electrons produced by geoneutrinos in the simulation.}
\label{fig:GeoDet}
\end{figure}

From the simulation, we can determine the average energy scale, {\it i.e.}, the number of PE per MeV, and
with that, the total number of detected PEs for each event is scaled to the detected energy.
The detected-energy spectra of solar- and geo-neutrinos are shown in Figure~\ref{fig:SolarDet} and Figure~\ref{fig:GeoDet}, respectively.

{We use a weighted-center method to reconstruct the direction of the recoiling electrons, $\vec{R}$. The formula is}

{\begin{equation}
\vec{R} = \frac{1}{N_{\rm{PE}}}\sum_{i=1}^{N_{\rm{PE}}} \vec{r}_{i},
\label{eq:RRec}
\end{equation}}
where $\vec{r}_{i}$ is the direction of each photoelectron and $N_{\rm{PE}}$ is the number of photoelectrons.

We tried three groups of photons. In Case (1) we use all Cherenkov photons to study the best case and to understand the scattering of electrons in the liquid scintillator.
In Case (2) we apply the 20\% efficiency cut as described in Section~\ref{sec:Detector}.
We use this study to understand the results with Cherenkov photons only.
In Case (3), we tested a more realistic case, where the detection efficiency is considered, and the photoelectrons from the first two ns of scintillation radiation are introduced.

\begin{figure}[!h]
\centering
\includegraphics[width=0.45\textwidth]{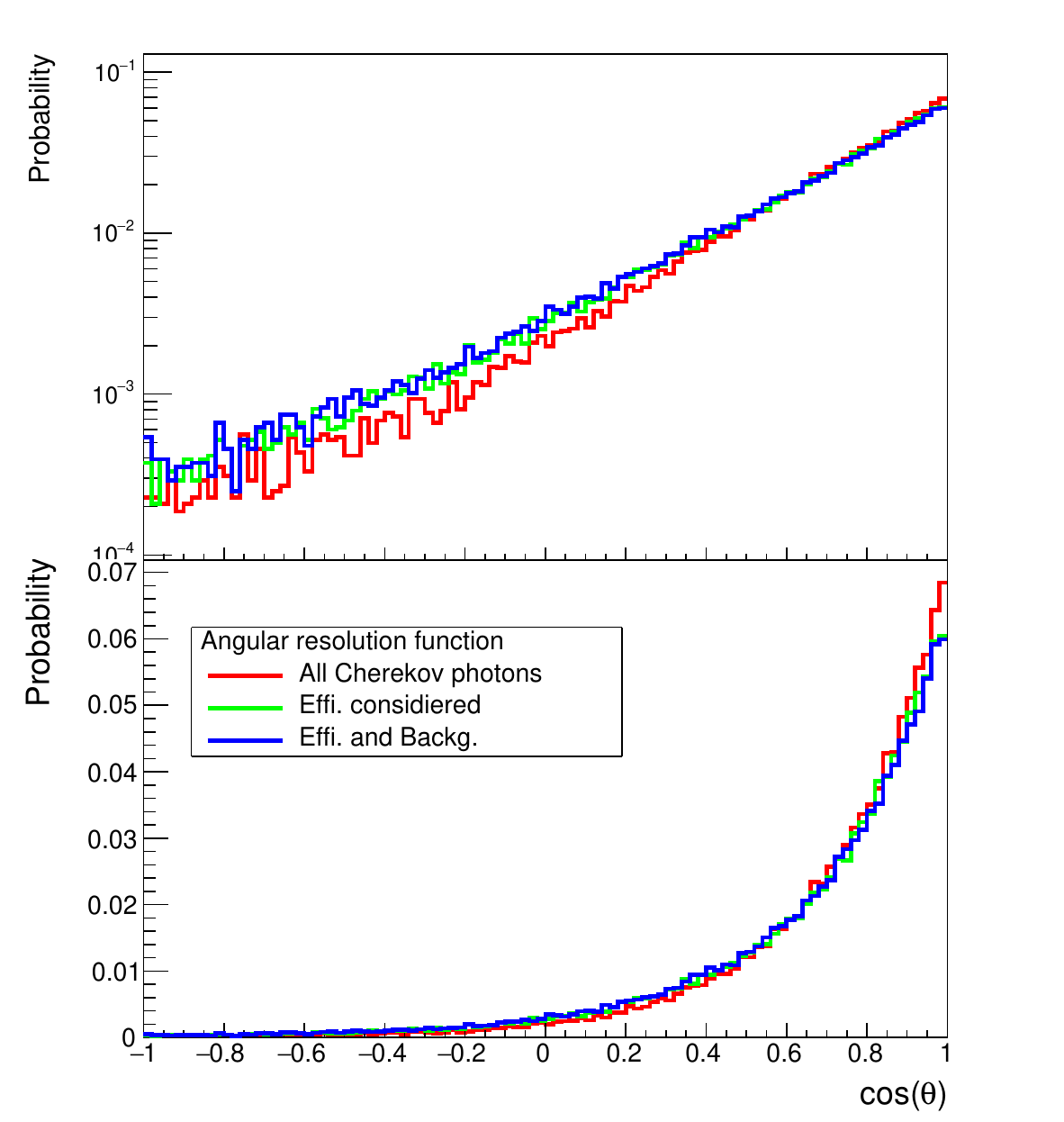}
\caption{Angular-response distribution of the reconstructed directions relative to the electrons' initial direction, where the electron kinetic energy is in the range [0.5, 2] MeV.
The upper and lower panels are the same, except that
the upper panel is plotted on a logarithmic scale to show the non-negligible negative component clearly.}
\label{fig:Resolution}
\includegraphics[width=0.45\textwidth]{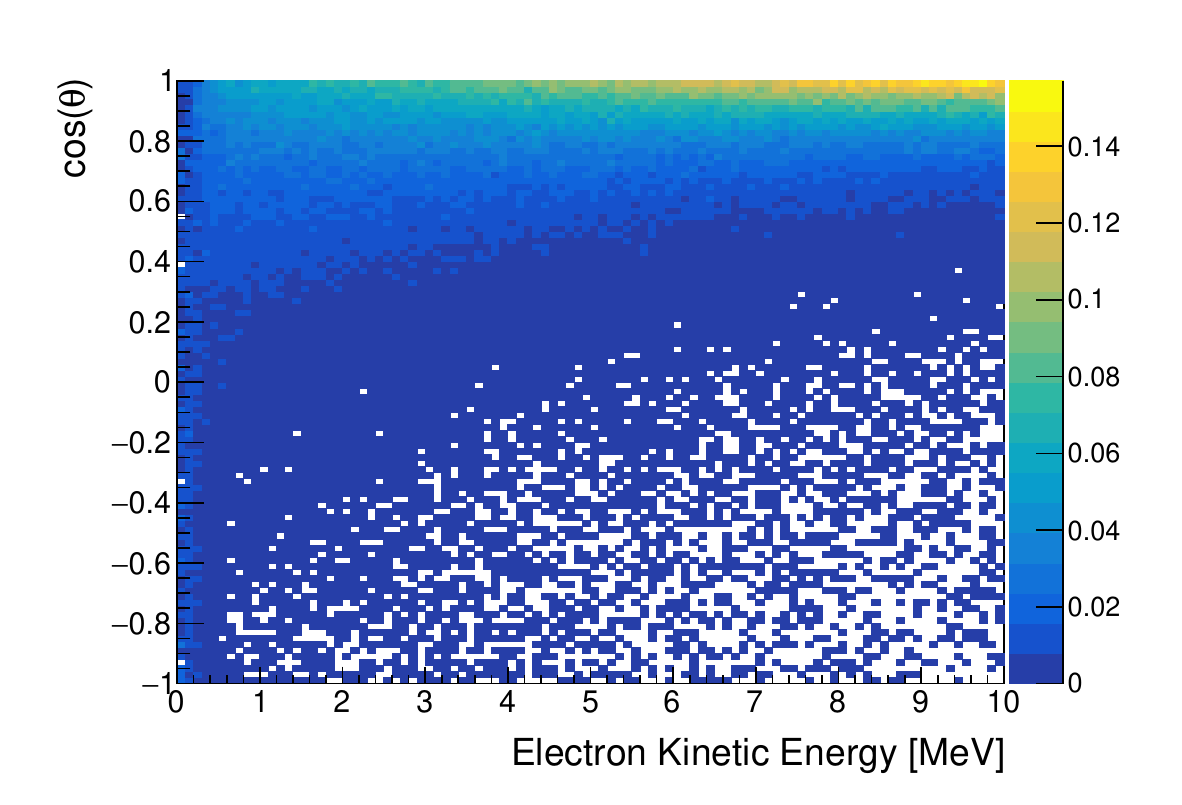}
\caption{Angular response versus energy. The projection on $\cos(\rm{\theta})$ for [0.5, 2] MeV electrons is shown in Figure~\ref{fig:Resolution}.}
\label{fig:ResolutionE}
\end{figure}

The angular response is plotted in Figure~\ref{fig:Resolution} for electrons with kinetic energies in the range of [0.5, 2] MeV,
where we show the cosine of the angle between the reconstructed direction and the initial electron direction for all three cases.
For case (1), the angular resolution with 99\% coverage is 116 degrees, and it is 124 and 125 degrees for the second and the third cases, respectively.
Comparing Case (1) with (2), which includes the 20\% efficiency cut, the latter doesn't show much degradation of the reconstructed angular distribution.
The dominant factor affecting the performance of directional reconstruction is electron scattering in the liquid scintillator.
After further introducing the scintillation photons as background in Case (3), we find that the angular resolution is slightly worse than for Case (2).
In the rest of this paper, we focus on Case (3), which is the more realistic one.

The angular response as a function of energy is shown in Figure~\ref{fig:ResolutionE} for Case (3).
The resolution improves gradually with increasing energy.

\subsection{Signal Extraction}
To extract the geoneutrino signals,
we first determine an energy cut and a $\cos\theta_{\odot}$ cut and
then do a statistical subtraction to remove the solar-neutrino background.

\subsubsection{Determination of the Signal Region}
\label{sec:signalRegion}
\begin{figure}[h]
\centering
\includegraphics[width=0.45\textwidth]{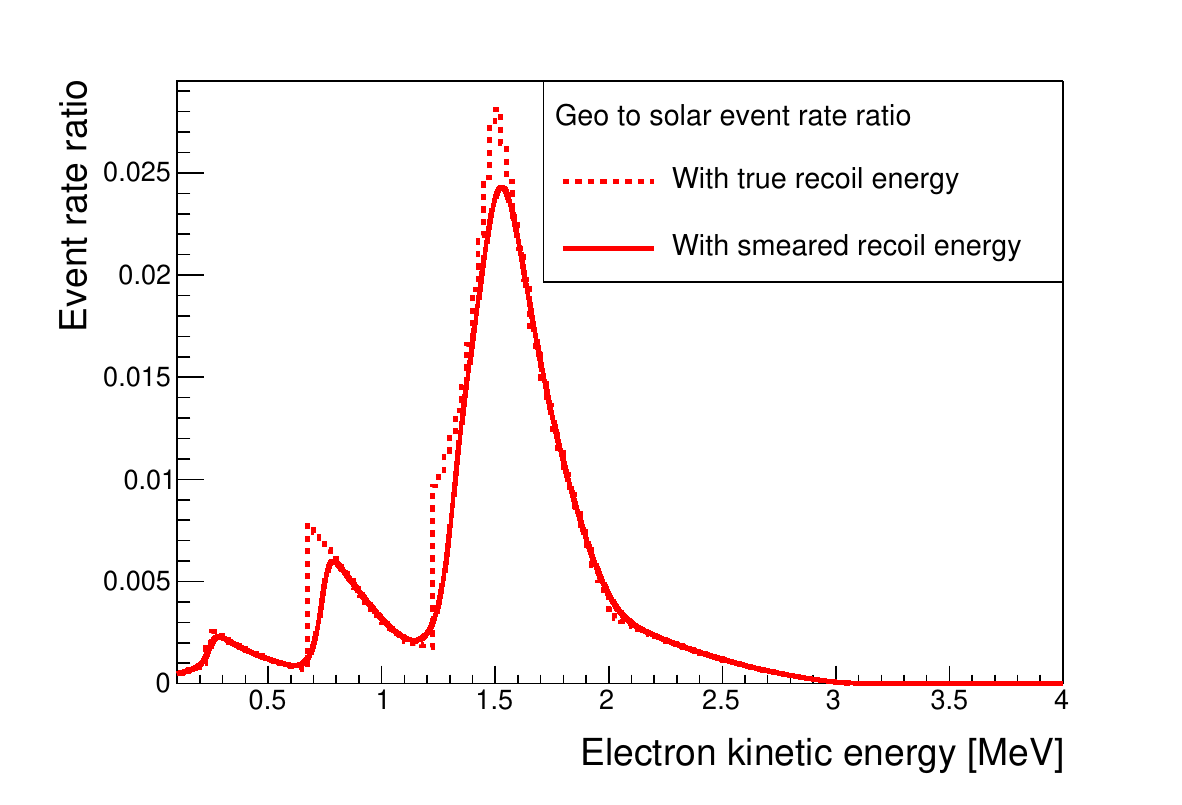}
\caption{Signal-to-background ratio (geo-to-solar-neutrino ratio) as a function of the kinetic energy of the electron.
}
\label{fig:Ratio}
\end{figure}

The event rate ratio of geo to solar neutrinos as a function of energy is shown in Figure~\ref{fig:Ratio}, according to which we define three observation windows.
One is the energy range [0.7, 2.3] MeV, where all \K, \Th, and \U\ geoneutrinos provide contributions.
The second is the range [0.7, 1.1] MeV, which is dominated by \K\ neutrinos, and the third is the range [1.1, 2.3] MeV, which is populated by \Th\ and \U\ neutrinos.
For the events below 0.7 MeV, like the $\nu_e$ component from \K\ decay, the directional reconstruction is rather poor,  and these events are not usable.
For all the three observation windows, the solar-neutrino background needs to be suppressed by a factor of 100-200 in order to enable us to extract the geoneutrino signals.
In the [0.7, 1.1] MeV window, the dominant solar-neutrino backgrounds are the $pep$, $^{13}$N, and $^{15}$O neutrinos,
while the $^{15}$O and $^{8}$B neutrinos are in the [1.1, 2.3] MeV window.

\begin{figure}[h]
\centering
\includegraphics[width=0.45\textwidth]{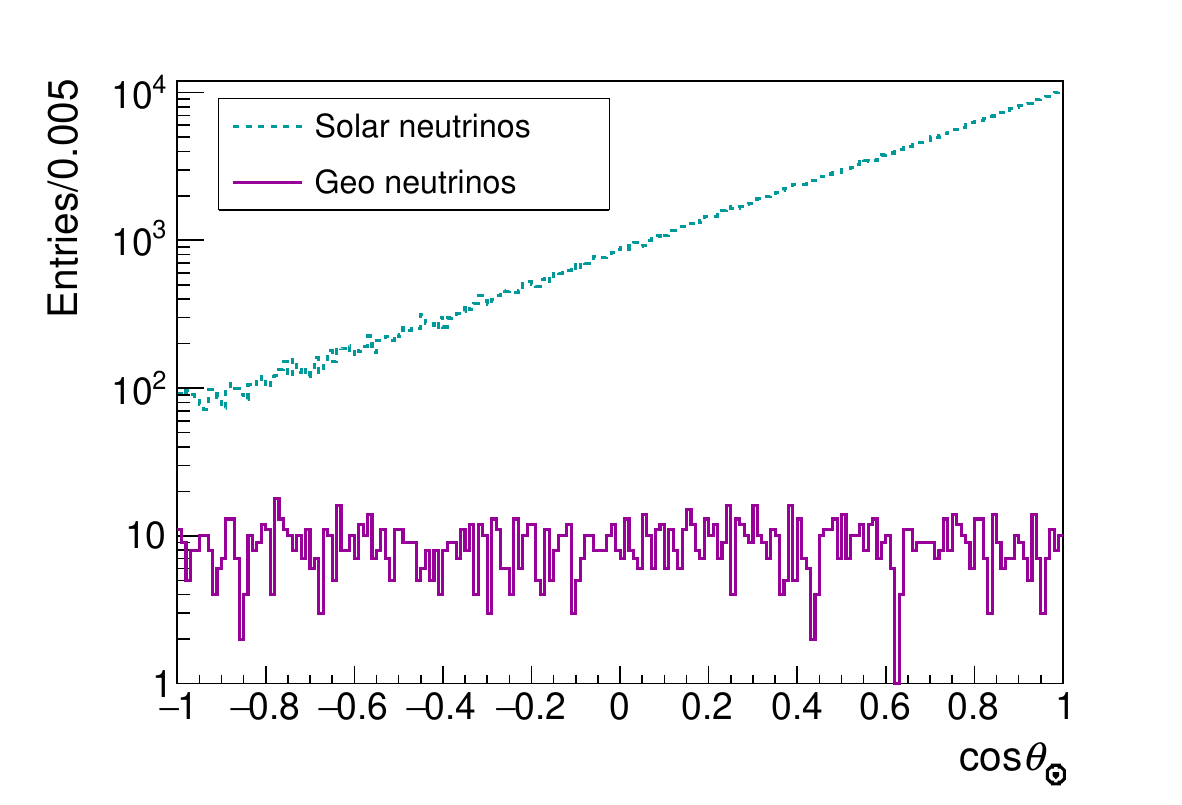}
\caption{The $\cos\theta_{\odot}$ distributions of the simulated solar and geo neutrinos, where the total statistics are for a 3 kt detector and a 20-year observation period, and a detected energy cut at [0.7, 2.3] MeV is applied.}
\label{fig:CosSunCut}
\end{figure}

We next determine the $\cos\theta_{\odot}$ cut.
After applying the detected-energy cut of [0.7, 2.3] MeV, the $\cos\theta_{\odot}$ of the remaining solar- and geo- neutrinos are both plotted in Figure~\ref{fig:CosSunCut}.
With a cut at $\cos\theta_{\odot}<-0.75$, the solar neutrinos are suppressed by a factor of 150, and the signal (geo) to background (solar) ratio is about 0.1, closer to unity than the other region.

\subsubsection{Signal Measurement}
{Using the data sample from our imagined experiment, the number of geoneutrino signals, $N_{geo}$, can be calculated by subtracting the solar-neutrino background:}

{\begin{equation}
N_{geo} = N_{can}-N_{bkg}\times \epsilon,
\label{eq:Sig}
\end{equation}}
where $N_{can}$ is the number of all candidates, $N_{bkg}$ is the background flux, {\it e.g.}, the solar neutrinos, and $\epsilon$ is the detection efficiency, including the energy-window cut and the $\cos\theta_{\odot}$ cut.

{The uncertainty $\sigma_{geo}$ in the geoneutrino counts is}

{\begin{equation}
\sigma_{geo} = \sqrt{\sigma_{candidate}^2 + N_{solar}^2\sigma_{\epsilon}^2 + \epsilon^2\sigma_{solar}^2},
\label{eq:Uncer}
\end{equation}}
where, $\sigma_{candidate}$ is the statistical uncertainty of the data sample,
$\sigma_{solar}$ is the solar-neutrino-flux uncertainty, and
$\sigma_{\epsilon}$ is the uncertainty in the efficiency.

For the solar-neutrino background, the pep and B8 neutrinos are dominant.
We expect several proposed experimental approaches, like Jinping~\cite{Jinping}, LENA~\cite{LENA}, THEIA~\cite{THEIA}, and~\cite{LAr}, will improve their uncertainty to a 1\% precision.
Calibration sources can be deployed to multiple locations of the detector~\cite{Reactor1}.
The calibration source can be a beta source enclosed in a small metal box with a small pinhole as a collimator. With enough statistics, a 1\% precision is expected.

From our experience, we assume that the detection-efficiency uncertainty, including the energy and the $\cos\theta_{\odot}$ cuts, can also reach 1\%.

For the \K\ energy window, we need in addition to subtract the \U\ and \Th\ geoneutrino components as backgrounds.
With the advantage of low reactor-neutrino background, the Jinping Neutrino Experiment can measure the total flux
of these neutrinos to better than 5\%.

\subsection{Sensitivity Curve}
{From the discussion above and the results for angular resolution and expected systematic uncertainties, we can now estimate the precision of the geoneutrino-flux measurement as a function of exposure. We express the sensitivity as}

\begin{figure}[!htb]
\centering
\includegraphics[width=0.45\textwidth]{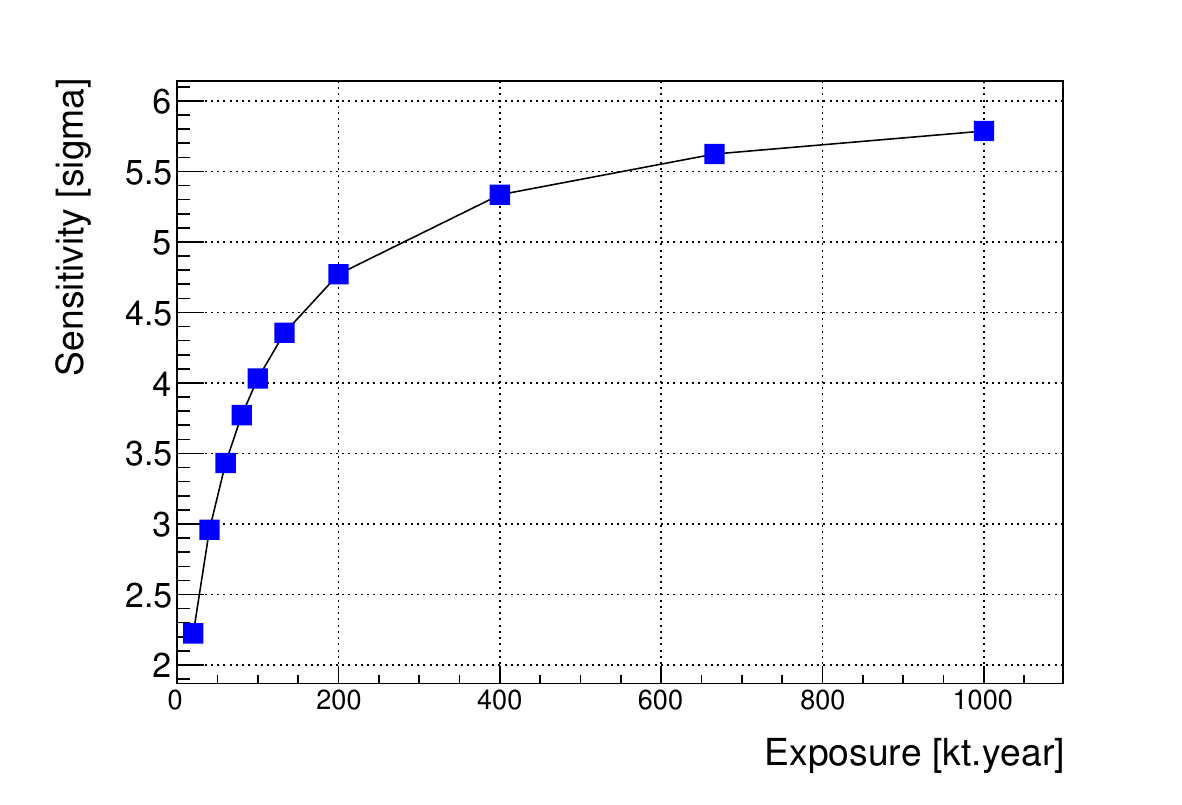}
\caption{Discovery sensitivity for all geoneutrinos as a function of exposure.}
\label{fig:AllSensitivity}
\includegraphics[width=0.45\textwidth]{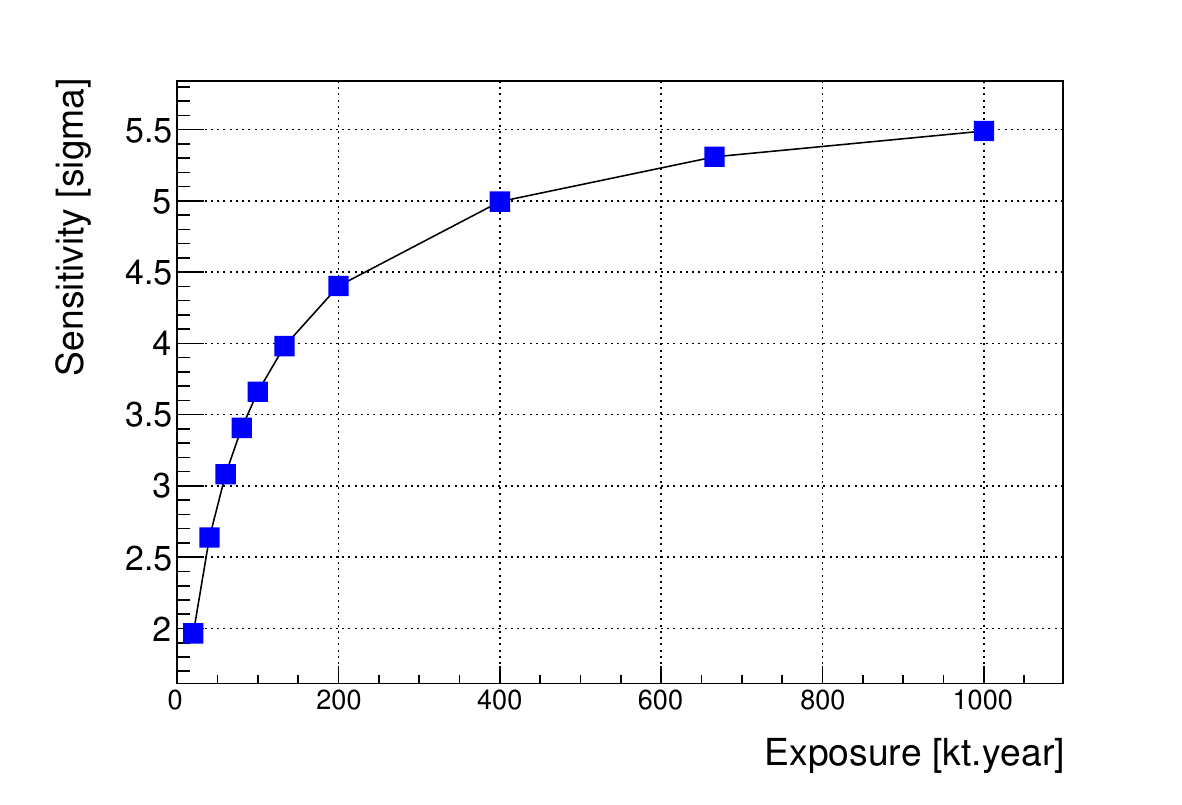}
\caption{Discovery sensitivity for \K\ geoneutrinos as a function of exposure.}
\label{fig:KSensitivity}
\includegraphics[width=0.45\textwidth]{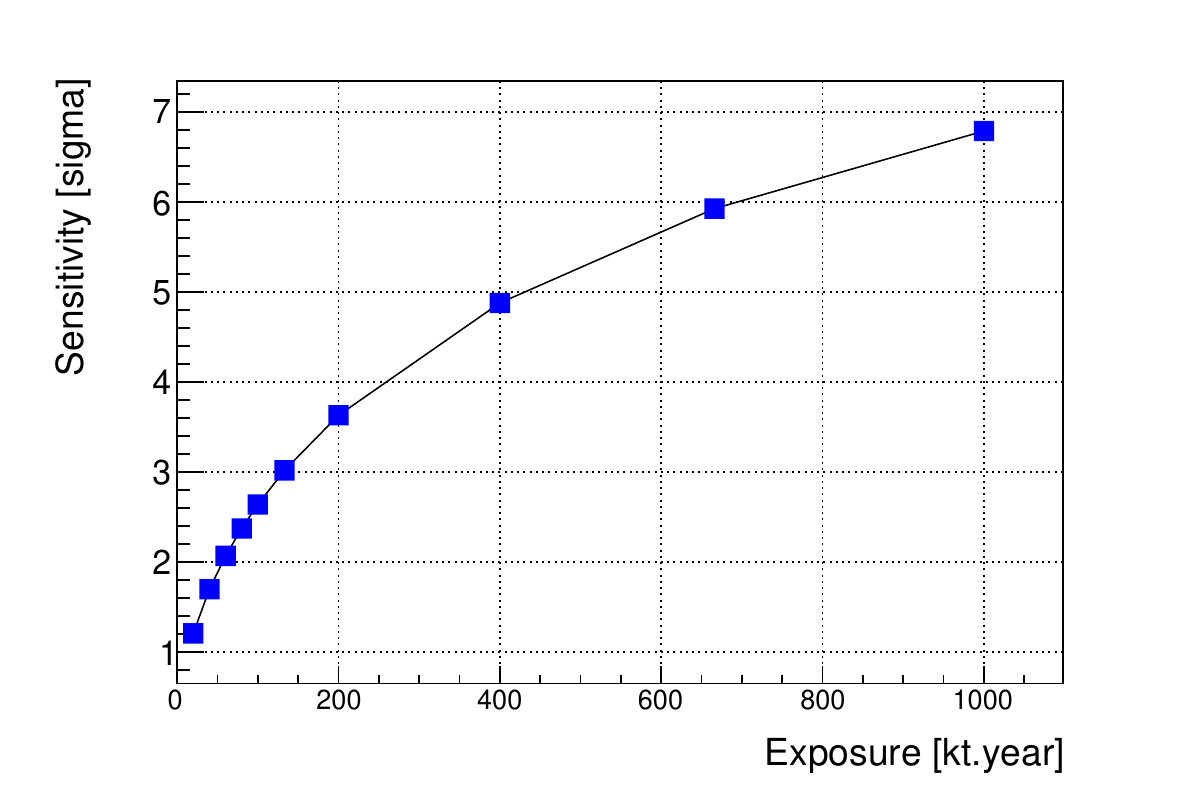}
\caption{Discovery sensitivity for \U\ and \Th\ genoneutrinos as a function of exposure.}
\label{fig:UThSensitivity}
\end{figure}

{\begin{equation}
\rm{sensitivity} = N_{geo}/\sigma_{geo},
\label{eq:Sens}
\end{equation}}
which gives the relative precision, or the deviation from the null assumption.
The study is performed for each of the three energy windows defined in Section~\ref{sec:signalRegion} above:
one for all geoneutrinos, [0.7, 2.3] MeV, another for the \K\ geoneutrinos, [0.7, 1.1] MeV, and the third for the \U\ and \Th\ components, [1.1, 2.3] MeV,
The results are shown in Figure~\ref{fig:AllSensitivity}, Figure~\ref{fig:KSensitivity}, and Figure~\ref{fig:UThSensitivity}, respectively.

Among these results, the most attractive is for the \K\ geoneutrinos, Figure~\ref{fig:KSensitivity}.
With a three-kiloton target mass and 20-year data-taking time, a 3-$\sigma$ observation is possible.
With a 20-kiloton detector, a 5-$\sigma$ observation is expected.

For the \Th\ and \U\ region, even with the better signal-to-background ratio shown in Figure~\ref{fig:Ratio},
the result is still limited by low statistics. So the expectation is worse than for the \K\ geoneutrinos.

\section{Discussion}
\label{sec:discussion}
The key aspects of this study are highlighted below, and the important properties of them will be discussed.
We generated the solar and geoneutrinos according to models, and propagated the neutrinos to a detector on the Earth's equator, taking into account neutrino oscillations.
Neutrino-electron elastic scattering is simulated using standard theoretical formulas.
The transport of the recoil electrons and the production of Cherenkov and scintillation photons are all handled by Geant4, using the customized light yield and rise and decay time constants of LAB.
Photoelectron detection is sampled according to a 20\% detection efficiency for a certain wavelength range.
The number of photoelectrons in each event is scaled to reconstruct the recoiling electron's kinetic energy.
A weighted-center method is applied to reconstruct the directions of the electrons.
With the reconstructed energy and directional distributions, we then determined the cuts required to extract geoneutrinos and to remove most of the solar neutrino background.
The remaining solar-neutrino background is subtracted statistically from the final sample.
We then scanned the exposure to see whether it is possible to discover geoneutrinos with this technique.
We elaborate on some features of this study below.

\subsection{Neutrino-Electron Scattering}
For this detection scheme, the directional reconstruction of the recoil electrons is crucial for the overall performance. The angular resolution governs the final signal-to-background ratio.
We find that the scattering of the electrons in the LAB has the primary effect on the resolution.
Reconstruction with a limited number of Cherenkov photoelectrons is only the secondary factor, as presented.
The density of the LAB is $0.87\times10^3$~kg/m$^3$. A further simulation study shows that
the angular resolution exhibits no significant improvement unless the density is close to the gaseous state.

\subsection{Slow Liquid Scintillator}
In this study, we assumed a 66.7\% detection efficiency, considering the PMT photocathode coverage, photon attenuation in the detector, and a 30\% quantum efficiency for photons in the range [300, 550] nm.
The detection efficiency is the most optimistic assumption in this entire study.
The scintillation emission spectrum of the pure LAB peaks at 340 nm~\cite{SlowLS}, which is close to the UV side and may suffer more absorption than we expected. The absorption is caused by the intrinsic absorption band of LAB and it cannot be resolved by purification.
With the addition of wavelength-shifting materials, the peak can be shifted into the visible range.
The absorption could be severer and some part of the Cherenkov light can be lost.
We hope that this investigation will stimulate further relevant slow-liquid-scintillator research,
for example, to search a new solvent and a new wavelength shifter.

\subsection{Other Background}
In this study, we included the critical solar-neutrino background, but other intrinsic or environmental radiative backgrounds should be considered too.
We take the situation of solar neutrino study at the Borexino experiment as an example~\cite{BXPhase1, BXLow} to explain our expectations.
The radioactive $^{10}\rm{C}$ and $^{11}\rm{C}$ background are induced by cosmic-ray muons.
At a deeper site, like the Jinping underground laboratory~\cite{Jinping}, these backgrounds will be suppressed by a factor of 100 or more and become negligible.
External photons affect the signal extraction, for example from $^{208}\rm{Tl}$, and this can be avoided by a tighter fiducial volume cut.
For the internal background, the decay products of the U and Th with secular equilibrium are not significant.
One thing worth special attention is the $^{210}\rm{Bi}$ background. After a few rounds of liquid scintillator purification with distillation, gas and water stripping and long term monitoring, the remaining $^{210}\rm{Bi}$ seems originally from radon gas absorption on detector inner surface and is leaching out by radon's daughter nuclei $^{210}\rm{Pb}$. Good progress has been made by suppressing the thermal convection of the liquid scintillator~\cite{BXLow2, BXLow3}. Surface cleaning was also mentioned to suppress the initial radon contamination. These effects should be considered later in developing a more realistic experimental design.

Reactor-neutrino backgrounds can be avoided by selecting an experimental site far away from commercial reactors, like the Jinping underground laboratory~\cite{Jinping, Linyan, Bill}. Reactor-neutrino fluxes also can be well constrained to be better than 6\%~\cite{Huber, Mueller, Reactor1, Reactor2}. The reactor-neutrino background can also be measured in-situ, so this is not a critical issue.

\subsection{Mantle Neutrinos}
Knowledge of mantle neutrinos also is necessary. However, it is only about 30\% of the total geoneutrino flux if the detector is placed on a continental site, while the rest originates from the crust. Given the current sensitivity in measuring the total flux and the current angular resolution, we did not pursue this issue further.

\section{Conclusion}
K element is volatile and its concentration in the Earth is not in balance with the refractory U and Th elements. A measurement of the K element in the Earth is of interest to understand the chemical evolution of the Earth.  The detection of \K\ neutrinos may lead to new knowledge of the Earth. Previously only U and Th geoneutrinos can be detected with the inverse beta process with a 1.8 MeV threshold. \K\ geoneutrinos are hard to discover for its low energy and high solar neutrino background. In this work, we found that liquid scintillator Cherenkov neutrino detectors can be used to detect the \K\ geoneutrinos. Liquid scintillator Cherenkov detectors feature both energy and direction measurements for charged particles. With the elastic scattering process of neutrinos with electrons, \K\ geoneutrinos can be detected without any intrinsic physical threshold. With the directionality, the dominant intrinsic background originated from solar neutrinos in common liquid scintillator detectors can be suppressed. With the studies of MeV electrons of Geant4 simulation, quantum and detection efficiency, and Cherenkov direction reconstruction, it is found that we can detect \K\ energy geoneutrinos with 3 standard deviations with a kilo-ton scale detector.
In this study, the setting of parameters is on the optimistic side, but we found that this technology is worth further development.

\section{Acknowledgement}
This work is supported in part by
the National Natural Science Foundation of China (Nos.~11620101004, 11235006, 11475093),
the Ministry of Science and Technology of China (no. 2018YFA0404102),
the Key Laboratory of Particle \& Radiation Imaging (Tsinghua University), and the CAS Center for Excellence in Particle Physics (CCEPP).

\appendix
\section{Solar-Neutrino Generation}
\label{apx:Solar}
We used the Standard Solar Model to provide the energy sampling of solar neutrinos.
Reference~\cite{JBHomepage} gives the neutrino-energy spectra of all solar neutrinos.
We used the neutrino-flux predictions on Earth, with the high-metallicity assumption from~\cite{MetalProb2} as normalization.
The characteristic energies and fluxes are summarized in Table~\ref{tab:solarFlux}, and the neutrino
energy spectra are shown in Figure~\ref{fig:SolarNu}.

\begin{table}[h]
\begin{center}
\caption{The characteristic energies and total fluxes of solar neutrinos.
The quantity E$_{\rm{Max}}$ is the maximum energy for continuous spectra and
E$_{\rm{Line}}$ is for discrete lines.
\label{tab:solarFlux}}
\footnotesize
\begin{tabular}{lcc}
\hline\hline
            & E$_{\rm{Max}}$ or E$_{\rm{Line}}$    & Flux                              \\
            &   [MeV]      & [$\times10^{10}$s$^{-1}$cm$^{-2}$]   \\ \hline
$pp$\ \     & 0.42 MeV     & $5.98(1\pm0.006)$                \\
$^7$Be\ \   & 0.38 MeV     & $0.053(1\pm0.07)$                \\
            & 0.86 MeV     & $0.447(1\pm0.07)$                \\
$pep$\ \    & 1.45 MeV     & $0.0144(1\pm0.012)$              \\
$^{13}$N\ \ & 1.19 MeV     & $0.0296(1\pm0.14)$               \\
$^{15}$O\ \ & 1.73 MeV     & $0.0223(1\pm0.15)$               \\
$^{17}$F\ \ & 1.74 MeV     & $5.52\times10^{-4}(1\pm0.17)$    \\
$^8$B\ \    & 15.8 MeV     & $5.58\times10^{-4}(1\pm0.14)$    \\
$hep$\ \    & 18.5 MeV     & $8.04\times10^{-7}(1\pm0.30)$    \\\hline\hline
\end{tabular}
\end{center}
\end{table}

Solar neutrinos are generated as pure electron neutrinos. Taking into account the oscillation between different neutrino flavors during transit in the Sun~\cite{MSW-W, MSW-MS}, the survival probability of electron neutrinos is~\cite{SPark, WHaxton}:

\begin{equation}
P^{\odot}_{ee}=\cos^4\theta_{13}(\frac{1}{2}+\frac{1}{2}\cos2\theta^M_{12}\cos2\theta_{12}),
\label{eq:Pee}
\end{equation}
where $\sin^2\theta_{12}$=0.307, $\sin^2\theta_{13}$=0.0241, and $\theta^M_{12}$ is the revised matter oscillation angle~\cite{MSW-W, MSW-MS, SPark, WHaxton}, which is neutrino energy and electron number density dependent~\cite{n_e}.

The appearance probability of $\nu_{\mu}$ or $\nu_{\tau}$ is

\begin{equation}
P^{\odot}_{e\mu(\tau)}=1-P^{\odot}_{ee}.
\end{equation}
The probability $P^{\odot}_{ee}$ ranges from 0.3 to 0.6. We did not consider neutrino oscillations in the Earth, because the change in probability is less than 5\%, which is insignificant for our study.

\section{Geoneutrino Generation}
\label{apx:Geo}
\begin{figure}[h]
\centering
\includegraphics[width=0.45\textwidth]{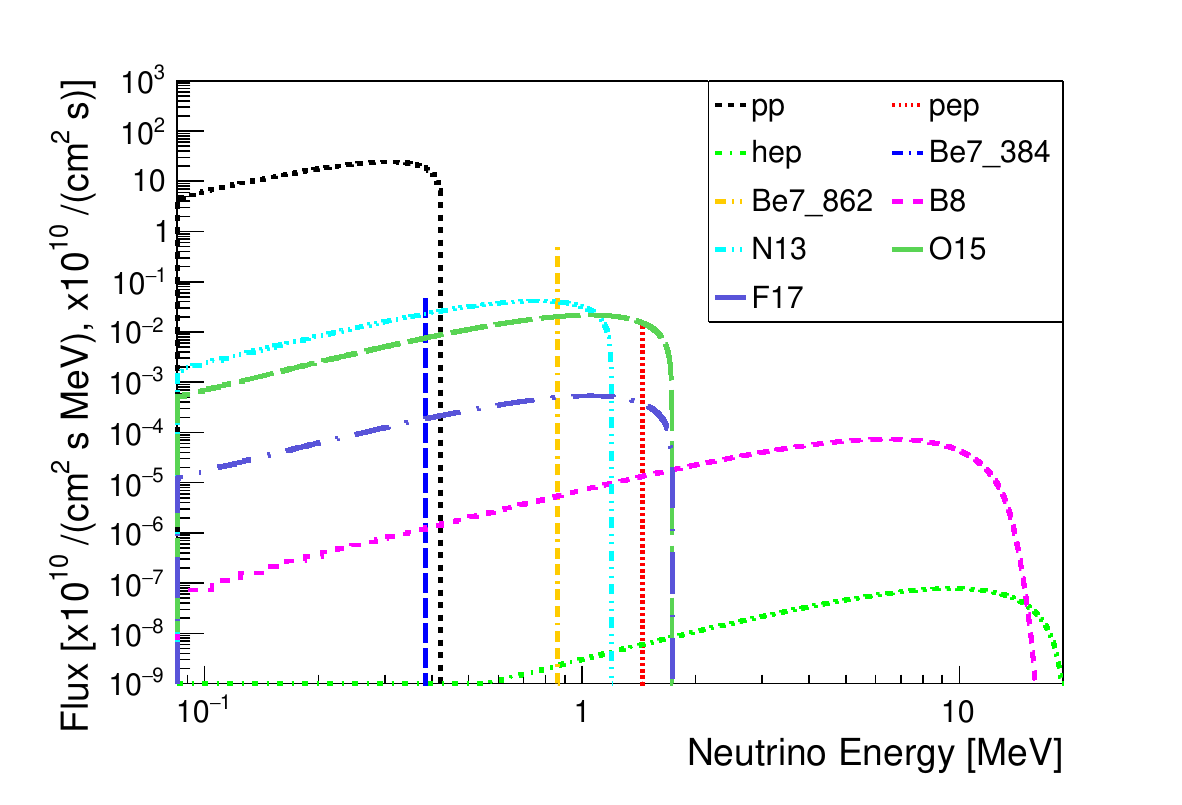}
\caption{Predicted non-oscillating solar electron-neutrino energy spectra on the Earth,
where the unit for the continuous spectra is $\rm{10^{10}/(cm^2\ s\ MeV}$, and for the discrete lines is
$\rm{10^{10}/(cm^2\ s)}$.}
\label{fig:SolarNu}
\includegraphics[width=0.45\textwidth]{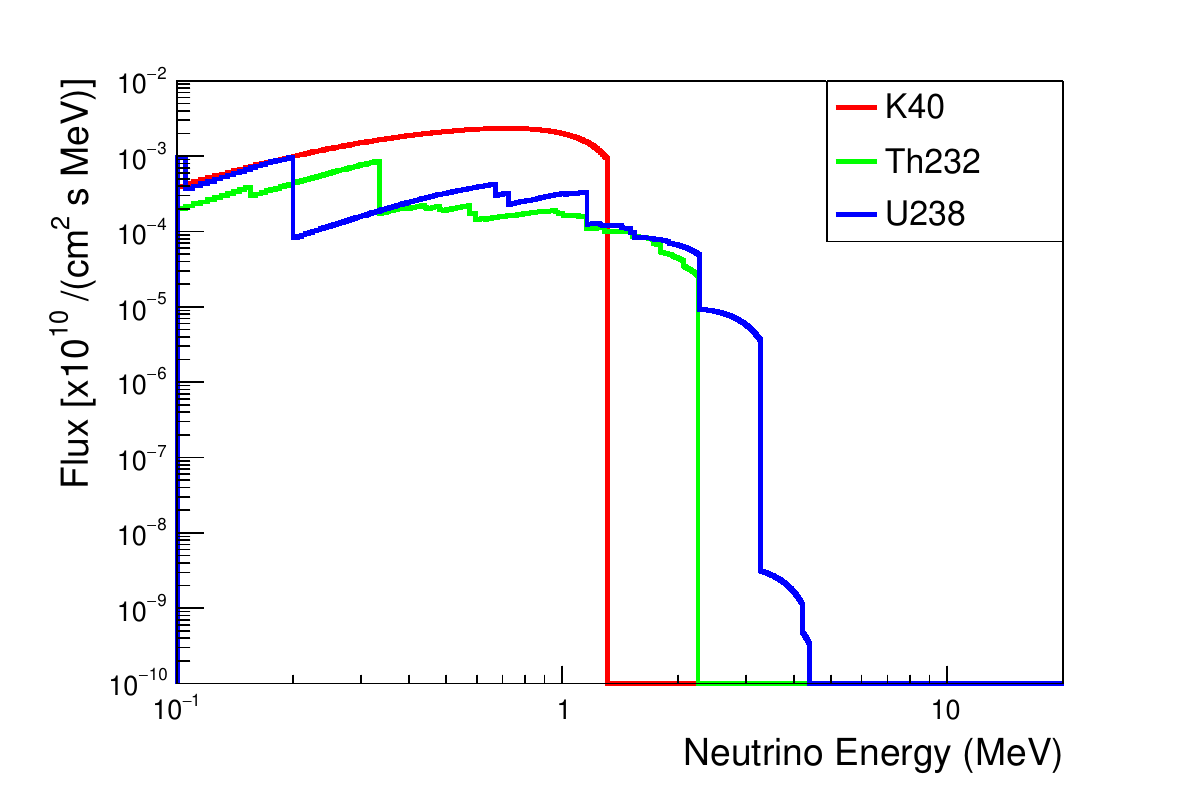}
\caption{Predicted non-oscillating geo electron-antineutrino energy spectra on the Earth's surface.}
\label{fig:GeoNu}
\end{figure}

There are three dominant heat-producing isotopes in the Earth: \U, \Th, and \K. Neutrinos are produced in their decay chains or direct decays:

\begin{align}
^{238}\rm{U}  \rightarrow ^{206}\rm{Pb} + 8\alpha + 6e^- + 6\bar\nu_e + 51.698~MeV,\\
^{232}\rm{Th} \rightarrow ^{207}\rm{Pb} + 7\alpha + 4e^- + 4\bar\nu_e + 46.402~MeV,\\
^{40}\rm{K} \rightarrow ^{40}\rm{Ca} + e^- + 4\bar\nu_e + 1.311~MeV~(89.3\%),\\
^{40}\rm{K} + e^- \rightarrow ^{40}\rm{Ar} + \nu_e + 1.505~MeV~(10.7\%).
\end{align}
Electron-antineutrinos are dominant, but 10.7\% of the \K\ decays occur through electron capture, producing electron-neutrinos with a maximum energy of 0.482 MeV.
This decay branch is ignored because it is hard to distinguish using the method proposed in this study.
The characteristic energies of these neutrinos are listed in Table~\ref{tab:GeoFlux}, and their
energy spectra can be seen in Figure.~\ref{fig:GeoNu}.
\begin{table}[h]
\begin{center}
\caption{The characteristic energies and total fluxes of geoneutrinos.
The quantity E$_{Max}$ is the maximum energy for continuous spectra.
\label{tab:GeoFlux}}
\footnotesize
\begin{tabular}{lccc}
\hline\hline
                         & Isotope& E$_{Max}$    & Flux                                      \\
                         &        &   [MeV]      & [$\times10^{10}$s$^{-1}$cm$^{-2}$]        \\ \hline
\multirow{3}{1cm}{Crust} &  \K    & 1.31 MeV     & 0.00160               \\
                         &  \Th   & 2.26 MeV     & 0.00043               \\
                         &  \U    & 3.27 MeV     & 0.00047               \\
\multirow{3}{1cm}{Mantle}&  \K    & 1.31 MeV     & 0.00057               \\
                         &  \Th   & 2.26 MeV     & 0.00005               \\
                         &  \U    & 3.27 MeV     & 0.00008               \\\hline\hline
\end{tabular}
\end{center}
\end{table}

We used a layered Earth model to simulate the geoneutrinos.
In this model, the Earth is taken to consist of 3 layers: the core, mantle, and crust.
We assume that the mantle and crustal layers have uniform distributions of $^{40}$K, $^{232}$Th and $^{238}$U,
and that there is no radioactivity from the core.

The whole volume of the Earth is divided into many small cells, each of which has a coordinate $\vec{r}$. Electron antineutrinos are sampled from each cell.
The differential flux of electron antineutrinos from each cell to the surface neutrino detector at $\vec{d}$ can be written as~\cite{Bill, Linyan}:

\begin{equation}
d\phi(\vec{r})_{e} = \frac{X \lambda N_A}{\mu} n_{\nu} P_{ee}^\oplus \frac{A(\vec r) \rho(\vec{r})}{4\pi |\vec{r}-\vec{d}|^2}dv,
\end{equation}
where $X$ is the natural isotopic mole fraction of each isotope, $\lambda$ is the corresponding decay constant, $N_A$ is the Avogadro's number, $\mu$ is the atomic mole mass, $n_{\nu}$ is the number of neutrinos per decay, $P_{ee}^\oplus$ is the average survival probability, $A(\vec r)$ is the abundance of each element in kg/kg, $\rho(\vec r)$ is the local density at each location, and $|\vec{r}-\vec{d}|$ gives the distance from each cell $\vec{r}$ to our detector at $\vec{d}$.

We take the outer radii of the core, mantle and crust to be 3480, 6321, and 6371 km~\cite{Earth1}, respectively,
with the corresponding densities to be 11.3, 5.0, and 3.0 g/cm$^3$.
The element abundance values of K, Th and U are set to match the integrated flux predictions, as in~\cite{Bill, Linyan}, and the values are given in Table~\ref{tab:abundance}.
This simplified layered model is not as sophisticated as that in reference~\cite{Crust1.0}, but it is sufficient for our demonstration purposes.
The rest of the parameters are taken from reference~\cite{Bill}.
The total fluxes of the predicted geoneutrinos are summarized in Table~\ref{tab:GeoFlux}.
The non-oscillating neutrino spectra of $^{40}$K, $^{232}$Th and $^{238}$U at the detection site are shown in Figure.~\ref{fig:GeoNu}.

The geoneutrino oscillation probability varies only by about 2\% in the energy range [0, 3.5] MeV~\cite{Linyan}, so it is treated as a constant, {\it i.e.}, $P_{ee}^\oplus=0.553$.
The appearance probability of the $\nu_{\mu}$ or $\nu_{\tau}$ components is

\begin{equation}
P^{\oplus}_{e\mu(\tau)}=1-P^{\oplus}_{ee}.
\end{equation}

\begin{table}
\begin{center}
\caption{Element abundance of K, Th and U in the mantle and crust used for this study.
\label{tab:abundance}}
\footnotesize
\begin{tabular}{lccc}
\hline\hline
            & K [kg/kg]           & Th [kg/kg]          &  U [kg/kg]             \\ \hline
Crust       & $1.16\times10^{-2}$ & $5.25\times10^{-6}$ &  $1.35\times10^{-6}$   \\
Mantle      & $152\times10^{-6}$  & $21.9\times10^{-9}$ & $8.0\times10^{-9}$   \\\hline\hline
\end{tabular}
\end{center}
\end{table}

\section{Neutrino-Electron Scattering}
\label{apx:Scat}
\begin{figure}[h]
\centering
\includegraphics[width=0.45\textwidth]{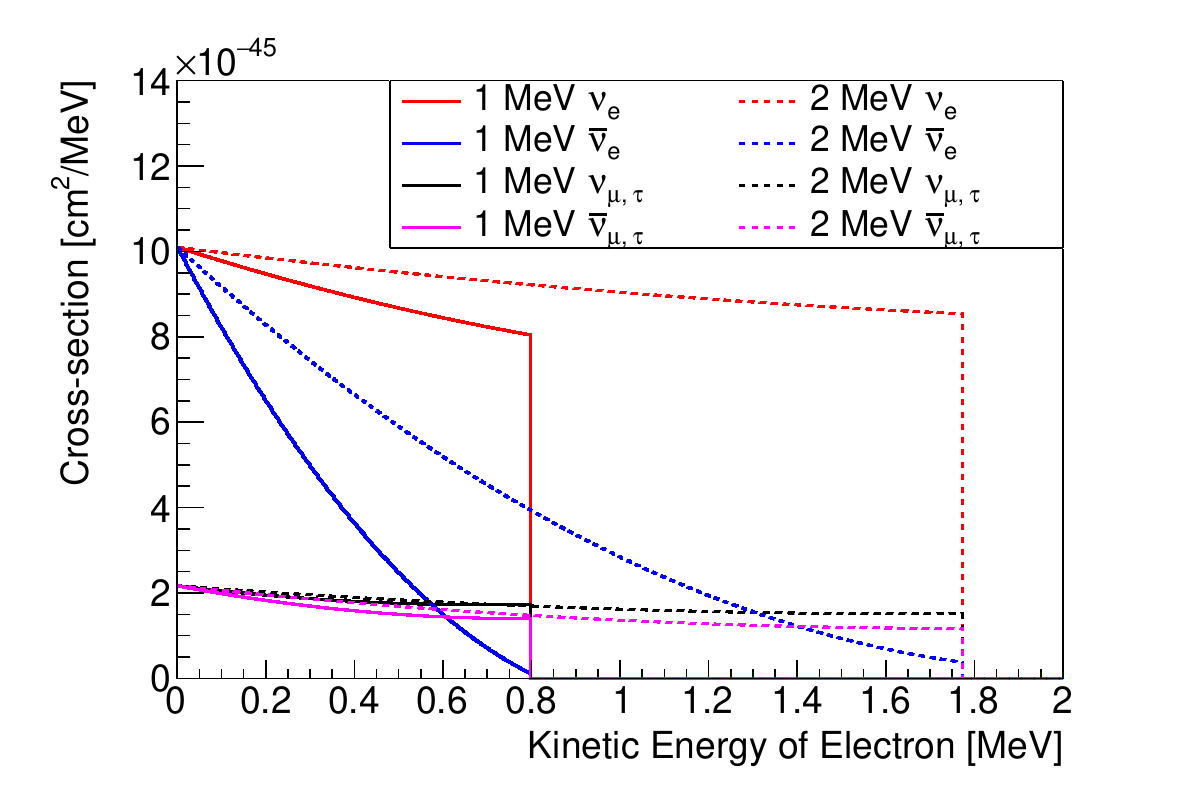}
\caption{Neutrino-electron-scattering differential cross-section for $\nu_e$, $\nu_{\mu,\tau}$, $\bar\nu_e$, and $\bar\nu_{\mu,\tau}$ at neutrino energies 1 and 2 MeV.}
\label{fig:CrossSection}
\end{figure}

The differential scattering cross-sections for neutrinos of energy $E_{\nu}$ and recoil electrons with kinetic energy $T_e$ can be written, {\it e.g.} in~\cite{NueXSec}, as:

\begin{equation}
\frac{d\sigma(E_{\nu},T_e)}{dT_e}=\frac{\sigma_0}{m_e}\left[g_1^2+g_2^2(1-\frac{T_e}{E_{\nu}})^2-g_1g_2\frac{m_eT_e}{E_{\nu}^2}\right],
\label{eq:CrossSection}
\end{equation}
where $m_e$ is the electron mass.
For $\nu_e$ and $\bar\nu_e$, $g_1$ and $g_2$ are:

\begin{equation}
\begin{split}
g_1^{(\nu_e)}&=g_2^{(\bar\nu_e)}=\frac{1}{2}+\sin^2\theta_W\simeq0.73,\\
g_2^{(\nu_e)}&=g_1^{(\bar\nu_e)}=\sin^2\theta_W\simeq0.23,
\end{split}
\end{equation}
where $\theta_W$ is the Weinberg angle, and for $\nu_{\mu,\tau}$, $g_1$ and $g_2$ are:

\begin{equation}
\begin{split}
g_1^{(\nu_{\mu,\tau})}&=g_2^{(\bar\nu_{\mu,\tau})}=-\frac{1}{2}+\sin^2\theta_W\simeq-0.27,\\
g_2^{(\nu_{\mu,\tau})}&=g_1^{(\bar\nu_{\mu,\tau})}=\sin^2\theta_W\simeq0.23.
\end{split}
\end{equation}
The constant $\sigma_0$ is

\begin{equation}
\sigma_0=\frac{2G_F^2m_e^2}{\pi}\simeq88.06\times10^{-46}\ \rm{cm}^2.
\end{equation}
The differential cross-section is shown in Figure~\ref{fig:CrossSection}. The antineutrinos cross-section is several times lower than for the neutrinos, and the recoil electrons produced by $\bar\nu_e$ tend to have lower kinetic energies.

With the above formulas, the distribution of the recoiling electrons' kinetic energy is calculated:

\begin{equation}
\frac{dN}{dT}=N_e\int
    [ \sum_{\nu} \frac{d\sigma(E_{\nu},T_e)}{dT_e} P_{e\nu} ]
    F(E_{\nu})dE_{\nu},
    \label{eq:sim}
\end{equation}
where $\frac{dN}{dT}$ is the number of scattered electrons $N$ per unit electron kinetic energy $T$,
$N_e$ is the number of target electrons,
the integral goes over all neutrino energies $E_{\nu}$,
the sum goes over all neutrino flavors $\nu$ which are $\nu_e$, $\nu_{\mu}$ $\nu_{\tau}$, $\bar\nu_e$, $\bar\nu_{\mu}$, and $\bar\nu_{\tau}$,
$\frac{d\sigma(E_{\nu},T_e)}{dT_e}$ is given by Equation (\ref{eq:CrossSection}),
$P_{e\nu}$ is the oscillation probability, and
$F(E_{\nu})$ is the flux of neutrinos.

\begin{figure}[!htb]
\centering
\includegraphics[width=0.45\textwidth]{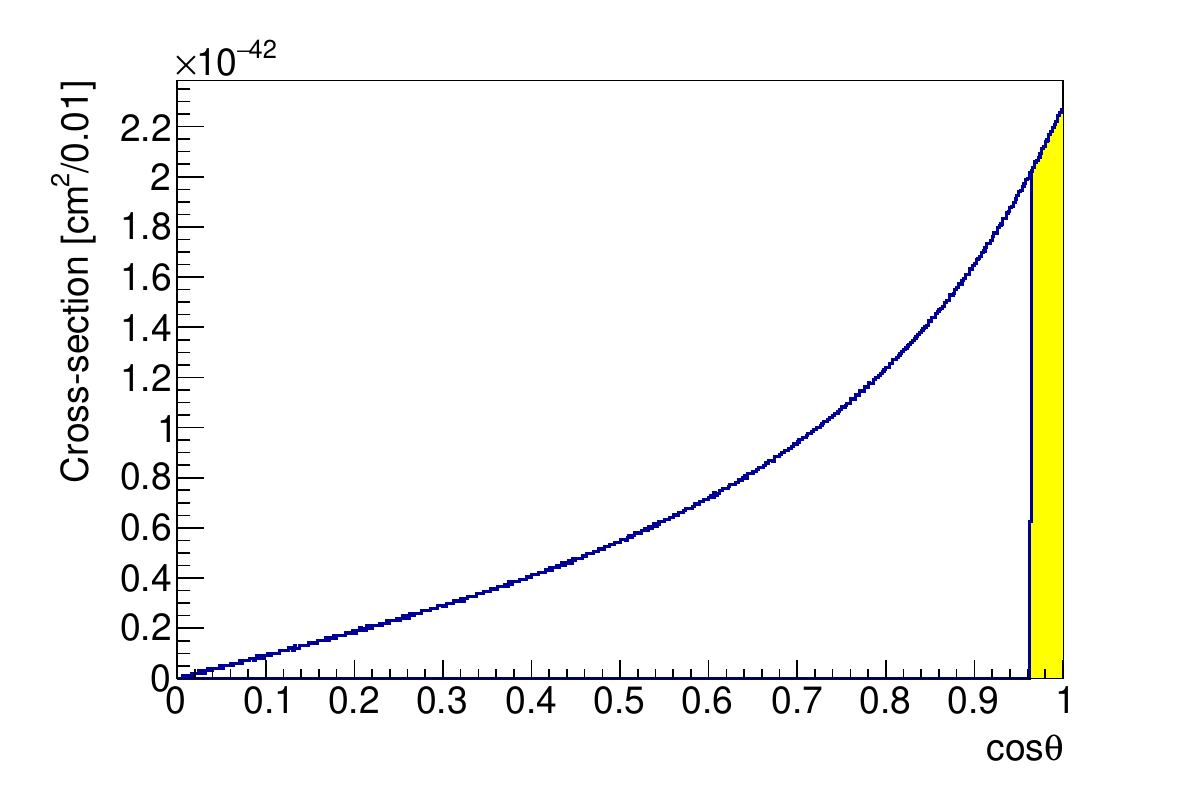}
\caption{The cosine distribution of the scattering angle between the initial neutrino direction and the recoiling electron. In this plot, we use 1 MeV $\nu_e$ as an example, and the shaded area show the result with a cut on the electron kinetic energy at 0.7 MeV.}
\label{fig:CosENu}
\end{figure}

With the condition of energy and momentum conservation, the cosine of the scattering angle between the initial neutrino direction and the scattered-electron direction can be determined from:

\begin{equation}
\cos\theta = \frac{1+m_e/E_{\nu}}{\sqrt{1+2m_e/T_e}}.
\end{equation}
The resulting $\cos\theta$ distribution is shown in Figure~\ref{fig:CosENu}. Note that although the directional correlation between the incoming neutrino and the recoil electron is weak at low energies, for example, at 1 MeV, with a cut on the kinetic energy of the recoil electron, the correlation still exists and can be employed. This feature is also shown in Figure~\ref{fig:CosENu}.

\begin{figure}[!h]
\centering
\includegraphics[width=0.45\textwidth]{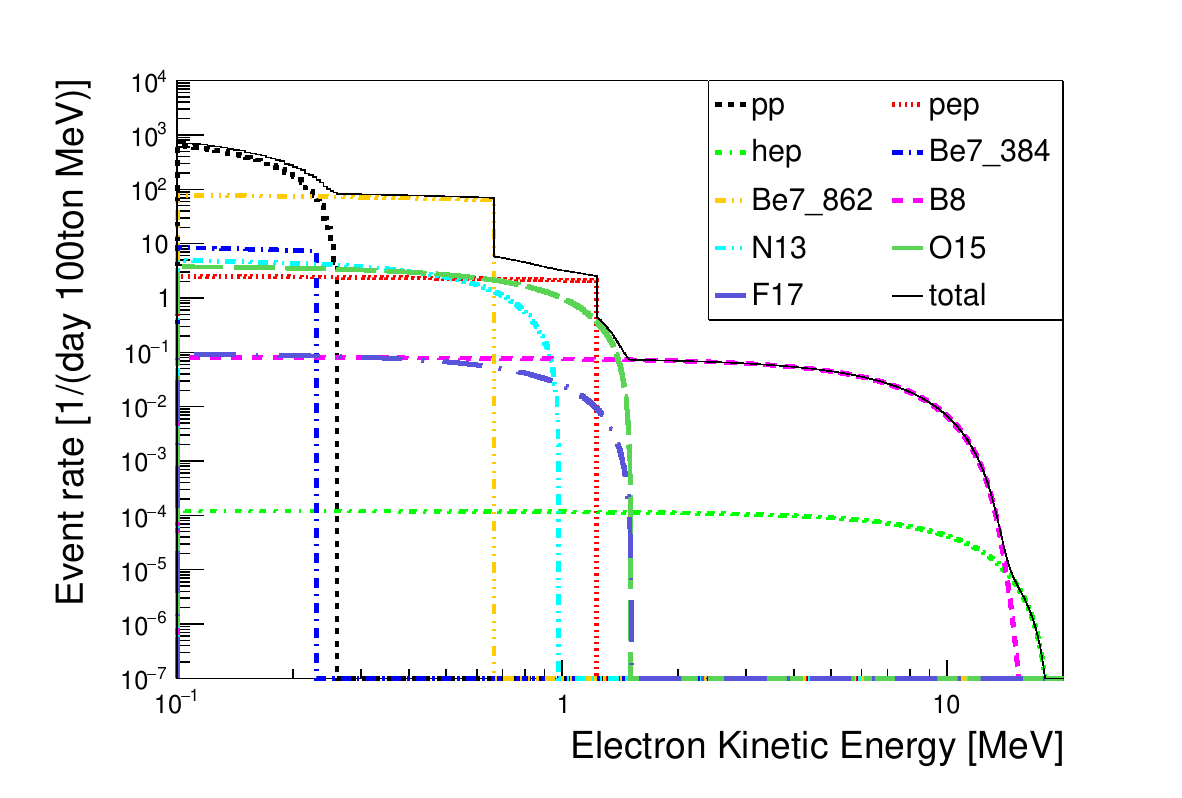}
\caption{Recoiling-electron kinetic-energy spectra from solar neutrinos.}
\label{fig:SolarTe}
\includegraphics[width=0.45\textwidth]{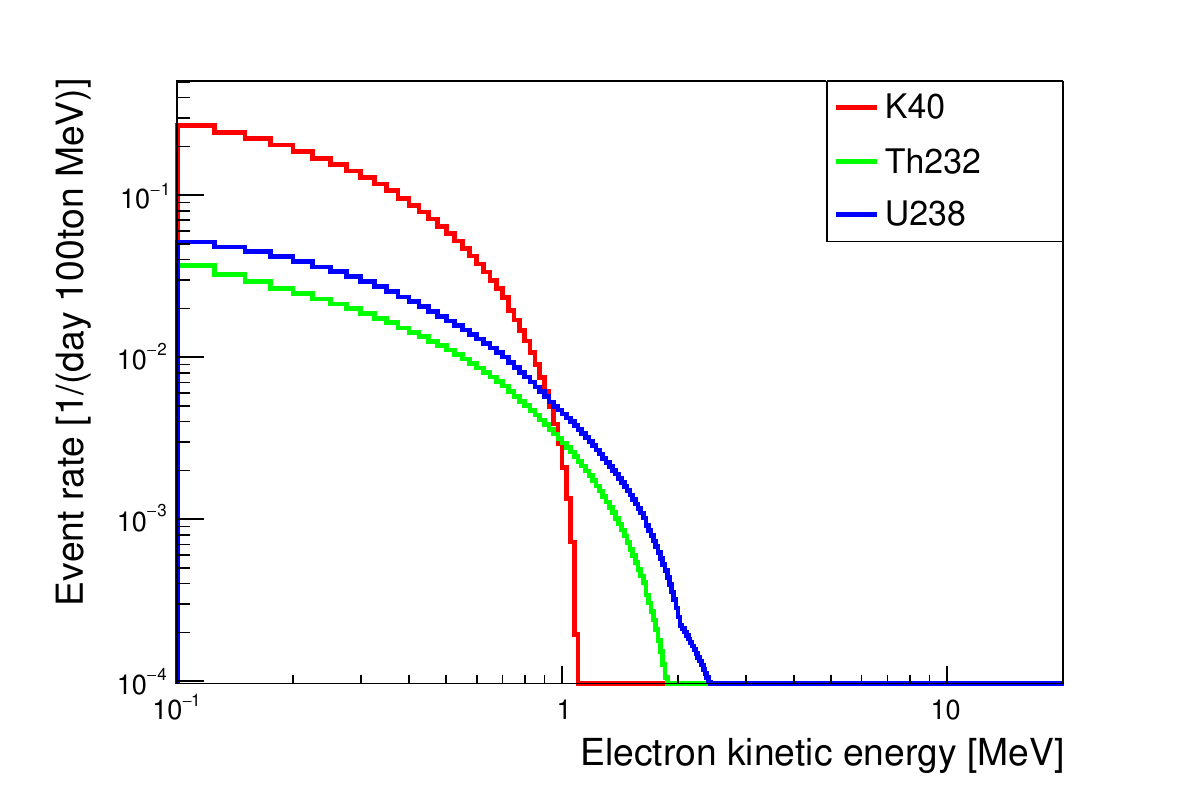}
\caption{Recoiling-electron kinetic-energy spectra from geo neutrinos.}
\label{fig:GeoTe}
\end{figure}

After considering neutrino oscillation and neutrino-electron scattering, the kinetic energy spectrum of recoiling electrons of solar- and geo-neutrinos are shown in Figure~\ref{fig:SolarTe}, and Figure~\ref{fig:GeoTe}, respectively.

\section*{References}
\bibliographystyle{elsarticle-num}
\bibliography{Geo}

\end{document}